\documentclass[10pt, conference, compsocconf]{IEEEtran}
\usepackage{enumerate}

\ifCLASSINFOpdf
  \usepackage[pdftex]{graphicx}
\else
  \usepackage[dvips]{graphicx}
\fi
\usepackage{array}
\usepackage{fixltx2e}

\usepackage{stfloats}
\begin{document}
%
\title{CRSTIP - An Assessment Scheme for Security Assessment Processes }


\author{\IEEEauthorblockN{
  Arthur-Jozsef Molnar}

\IEEEauthorblockA{
	Info World\\
	Bucharest, Romania\\
 arthur.molnar@infoworld.ro
}
\and
\IEEEauthorblockN{
  J\"urgen Gro\ss{}mann
}
\IEEEauthorblockA{
  Fraunhofer FOKUS\\
  Berlin, Germany\\
  juergen.grossmann@fokus.fraunhofer.de
}

}



\maketitle

\begin{abstract} Complex networked systems are an integral part of today's
support infrastructures. Due to their importance, these systems become more and
more the target for cyber-attacks, suffering a notable number of security
incidents. Also, they are subject to regulation by national and international
legislation. An operator of such an infrastructure or system is responsible for ensuring
its security and correct functioning in order to satisfy customers. In addition,
the entire process of risk and quality control needs to be efficient and
manageable. This short paper introduces the Compliance, Risk Assessment and
Security Testing Improvement Profiling (CRSTIP) scheme. CRSTIP is an evaluation
scheme that enables assessing the maturity of security assessment processes,
taking into consideration systematic use of formalisms, integration and
tool-support in the areas of compliance assessment, security risk assessment and
security testing. The paper describes the elements of the scheme and their
application to one of the case studies of the RASEN research project.
\end{abstract}


%
\IEEEpeerreviewmaketitle

\section{Introduction}

Researchers within the RASEN \footnote{The FP7 RASEN project,
http://www.rasenproject.eu} project develop methods dedicated to supporting
companies and organizations in undertaking risk analysis for large scale,
networked systems. These methods cover security risk assessments on different
levels of abstraction and from different perspectives. Compliance assessment
especially addresses compliance of products and processes for which regulations
are in effect. Security risk assessment deals with the concise assessment of
security threats, estimating the probabilities and consequences for a set of
technical or business related assets. Finally, security testing can be used to
examine the target under assessment, be it an organization or system for actual
weaknesses or vulnerabilities. While the industry demands integrative approaches
that cope with security as a whole, currently no established process exists that
sufficiently emphasizes the systematic integration of compliance assessment,
security risk assessment and security testing. Within the RASEN project we aim
to close this gap by developing an integrated security assessment framework
based on compliance assessment, security risk assessment and security
testing. The resulting framework will be evaluated using three industrial case
studies.

Currently, there exist a number of methods to evaluate the maturity and quality
of test and assessment processes. The most known representative is the Test
Process Improvement (TPI) and its successor TPI NEXT \cite{TPINext}. Both
schemes are trademarks of SOGETI \cite{SOGETI} and have been applied to assess
industrial  processes across the world. Another approach is the Test Maturity
Model (TMM) and its successor Test Maturity Model integration (TMMi)
\cite{TMMI}. However, both approaches emphasize on testing and do not
sufficiently cover the aspects of compliance assessment and risk assessment as
required to assess the RASEN approach.

\section{The CRSTIP approach for process evaluation} The CRSTIP (Compliance and
Risk Security Testing Improvement Profiling) evaluation scheme can be used to
assess the readiness level of an organization, process or system with regards to
four key areas: legal and compliance assessment, security risk assessment,
security testing and tool support and integration. CRSTIP was initially used to
assess the baseline of the RASEN use cases, that is their status quo before
applying the techniques and tools that are developed within the project. It has
been additionally used to express expectations regarding the progress within the
four key areas for each of the case studies during the project's life time. The
scheme will be used again in order to document the actual progress achived after
deploying RASEN methodology and tooling.

CRSTIP provides a simple, straightforward assessment with regards to the
target's current positioning within the CRSTIP key areas. The approach is based
on the general ideas of TMMi and TPI and previous work undertaken within the
ITEA2 DIAMONDS project \footnote{The ITEA2 DIAMONDS project,
  http://www.itea2-diamonds.org}, where it was limited to assess progress in
selected key areas of security testing \cite{ETSI_TR101582}.  For each of the
key areas we defined a performance scale with a four-level hierarchy that can be
used to evaluate security assessment processes with respect to
performance. Within each area, levels with a higher number represent an
improvement over lower levels. We plan to further refine CRSTIP within our
project in order to serve as liaison between project efforts and organizations
seeking to improve their standing in the key areas addressed by RASEN. This
paper details the CRSTIP key areas and levels, showing its initial application
to the Medipedia system.  The key areas and their levels are detailed in the
following subsections.

\subsection{Key area - Legal and compliance assessment} This key areas refers to
the overall process that is employed with the objective of adhering to the
requirements of laws, to industry and organizational standards and codes, to
principles of good governance and accepted community as well as to ethical
standards.  The overall process should support, to the extent possible, the
documentation of compliance with these laws, rules and norms. The levels of this
key area are:
\\
\textbf{Level 1: Ad-hoc.} The compliance assessment is unstructured, does not use a
defined compliance process, and compliance decisions are made primarily on an
event-driven basis.\\
\textbf{Level 2: Check list based.} The checklist-based compliance assessment uses a
checklist to answer a set of standard questions or to tick checkboxes.\\
\textbf{Level 3: Systematic.}  A systematic compliance assessment follows a structured and planned approach where there is a defined process and structured
documentation of compliance. Generally, the process involves the identification
of compliance requirements, evaluation of the compliance issues and taking
measures to ensure compliance.\\
\textbf{Level 4: Systematic and risk-driven.} A systematic and risk-driven compliance assessment involves a defined process for risk-driven compliance where
requirements are prioritized based on their risks. This approach is supported by a systematic documentation that enables the mapping of different risks and controls to relevant compliance requirements.

\subsection{Key area - Security risk assessment} Risk assessment is the overall
process of risk identification, estimation and evaluation. Risk identification
is the process of finding, recognizing and describing risks. This involves
identifying sources of risk, areas of impact and events, together with their
causes and potential consequences. Risk estimation is the process of
comprehending the nature of risk and determining its level. Finally, risk
evaluation is the process of comparing the results of risk estimation with risk
criteria to determine whether the magnitute of risk is acceptable. Risk
evaluation assists in decisions about risk treatment. The levels of this key
area are:
\\
\textbf{Level 1: Checklist.} Risk assessment mainly consists of answering a
sequence of questions or filling in a form.\\
\textbf{Level 2: Qualitative.} Risk assessment is based on qualitative risk
values. The value descriptions or distinctions are based on some quality or
characteristic rather than on some quantity or measured value. \\
\textbf{Level 3: Quantitative.} Risk assessment is based on quantitative
values. The values are based on some quantity or number, e.g. a measurement, rather than
on some quality. \\
\textbf{Level 4: Real time.} Risk assessment is done in real-time based on an underlying,
computerized monitoring-infrastructure.

\subsection{Key area - Security testing}
Security testing is used to empirically check software implementations with
respect to their security properties and resistance to attack. Functional
security testing is used to check the functionality, efficiency and availability
of security features of a dedicated test item. Security vulnerability testing
directly addresses the identification and discovery of system
vulnerabilities. It targets the identification of design flaws and
implementation faults that can harm the availability, confidentiality and
integrity of the test item. The levels of this key area are:
\\
\textbf{Level 1: Unstructured.}  Unstructured security testing is performed either by the development team or the testing team without planning or documentation. The tests are intended to be run only once, unless a defect is discovered. The testing is neither systematic nor planned. Defects found using this method may be harder to reproduce.\\ 
\textbf{Level 2: Planned.} Planned security testing is performed either by the
development team or the testing team after a structured test plan has been elaborated. A test plan documents the scope, approach, and resources that will
be used for testing.\\
\textbf{Level 3: Risk based.} Security tests are planned and executed, either by
the development team or by the testing team. The planning of security testing is done on the basis of the security risk assessment using impact estimations or likelihood values to focus the testing process.\\
\textbf{Level 4: Continuous risk based. } Continuous risk based security testing is a process of continuously monitoring and testing a system with respect to potential
vulnerabilities. Security risk analysis results are still used to focus the
security testing and optimize resource planning. Any evolution of the system, of
its environment or of the identified threats leads to updated security tests so
that vulnerabilities can be detected throughout the whole life cycle of the test
item. 

\subsection{Key area - Tool support and integration}
This key area describes the degree of tool support and integration available for
the above mentioned areas. Typically, tools work on their own data structures
that are well suited to the task which needs to be performed with or by the
tool. Tool integration is the ability of tools to cooperate by exchanging data
or sharing a common user interface. The levels of this key area are:
\\
\textbf{Level 1: None.} No tool support in any of the above mentioned key areas is available.\\
\textbf{Level 2: Stand-alone.} Tools are available for some of the previously mentioned key areas. However, the tools are not integrated thus they do not exchange data nor do they share the same user interface.\\
\textbf{Level 3: Partially integrated.} Tools are available for some of the above mentioned key areas. Tool integration is based on point-to-point coalitions between tools. Point-to-point coalitions are often used in small and ad-hoc environments but have problems when it comes to more tools and larger environments as they do not scale.\\ 
\textbf{Level 4: Integrated.} Tools are available for nearly all the key
areas. Tool integration is based on central integration platforms and
repositories that provide a common set of interfaces and data definitions to be exchanged.

\begin{figure}[h]
    \centering
    \includegraphics[width=\linewidth]{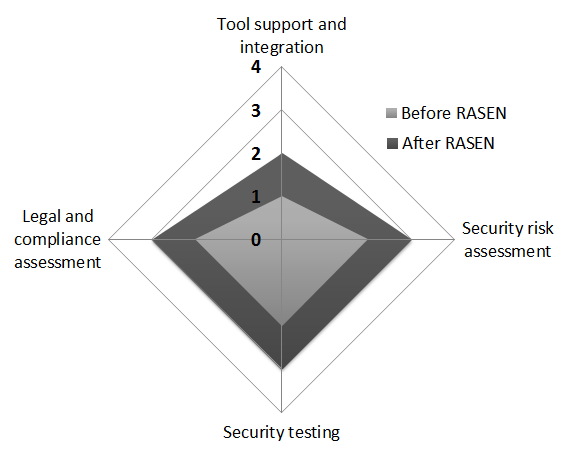}
    \caption{CRSTIP Evaluation of Medipedia}
    \label{fig:CRSTIPMedipedia}
\end{figure}

\section{The Medipedia case study } Medipedia\footnote{http://www.medipedia.ro/}
is an eHealth web portal developed by Info World that differentiates itself on
the market by allowing users to build and manage their personal electronic
healthcare record. As complex networked software system, Medipedia has over 36000
active users and must fulfil legal requirements with regards to processing
highly-sensitive personal data such as medical analyses results and diagnostic
history. As a case study system for RASEN, we have employed CRSTIP to Medipedia
in the following way: first, we evaluated the baseline, shown in Figure
\ref{fig:CRSTIPMedipedia}, as "Before RASEN". Then, based on preliminary project
results we estimate the benefits of implementing RASEN, shown on the same
figure.

As the system processes sensitive customer data, key areas already present
maturity. However, it is clear that a structured approach will benefit Medipedia
in virtually all of them. First of all, while the system is legally compliant, a
structured approach enables Info World to better prepare for upcoming
regulations such as the General Data Protection Regulation\cite{GDPR} and
facilitates cornering new markets having different regulations. Furthermore,
while the system undergoes planned security testing and periodical risk
assessment, there is no interplay between these activities. A structured risk
assessment process that enables Info World to guide testing and which can be
updated using test results facilitates bringing new features to market
faster. The final key area concerns software tooling, where Info World
recognizes the advantages supportive tooling would bring to its risk assessment
and testing processes. 

\section{Conclusion and Outlook}
CRSTIP was developed as an objective analysis and evaluation scheme of the
research and development within RASEN. Currently we have used it to assess the
case studies' baseline and to outline progress expectations for the end of the
project. We believe that in its current form, CRSTIP is a useful tool which
stakeholders can use to asses a target organization, process or product. More
so, as shown above, the scheme can be used to gain understanding about which
areas are most suitable for further investment and how the levels in the
different key areas relate or require each other.

Furthermore, we envision using CRSTIP as a dissemination tool for RASEN
technologies, as it allows identifying maturity levels with respect to key
security and compliance areas. Ideally, a concise description for each of the
key areas should be available that denote the techniques and tools that can be
used to drive the improvement as well as the requirements to other key areas
that are the precondition to improve from one level to the next. As future work,
our desire is to provide a web-based implementation where users are able to fill
in their assessment and obtain information regarding the requirements for moving
to the next level in their areas of interest.


\end{document}